# Spoken Language Identification Using Hybrid Feature Extraction Methods

Pawan Kumar, Astik Biswas, A .N. Mishra and Mahesh Chandra

**Abstract**—This paper introduces and motivates the use of hybrid robust feature extraction technique for spoken language identification (LID) system. The speech recognizers use a parametric form of a signal to get the most important distinguishable features of speech signal for recognition task. In this paper Mel-frequency cepstral coefficients (MFCC), Perceptual linear prediction coefficients (PLP) along with two hybrid features are used for language Identification. Two hybrid features, Bark Frequency Cepstral Coefficients (BFCC) and Revised Perceptual Linear Prediction Coefficients (RPLP) were obtained from combination of MFCC and PLP. Two different classifiers, Vector Quantization (VQ) with Dynamic Time Warping (DTW) and Gaussian Mixture Model (GMM) were used for classification. The experiment shows better identification rate using hybrid feature extraction techniques compared to conventional feature extraction methods.BFCC has shown better performance than MFCC with both classifiers. RPLP along with GMM has shown best identification performance among all feature extraction techniques.

**Index Terms**—Bark Frequency Cepstral Coefficient, Dynamic Time Warping, Language Identification, Gaussian Mixture Model, Mel Frequency Cepstral Coefficient, Perceptual Linear Prediction, Revised Perceptual Linear Prediction, Vector Quantization

——————————— ◆ ———————————

## 1 INTRODUCTION

Now Language Identification (LID) systems is an integral part of telephone and speech input computer networks which provide services in many languages. Automatic language identification (language ID for short) is the problem of identifying the language being spoken from a sample of speech by a speaker.

As with speech recognition, humans are the most accurate language identification systems in the world today. Within seconds of hearing speech, people are able to determine whether it is a language they know. If it is a language with which they are not familiar, they often can make subjective judgments as to its similarity to a language they know, e.g., "sounds like Tamil". A LID [1] system can be used to pre-sort the callers into the language they speak, so that the required service will be provided in the language appropriate to the talker. Examples of these LID services includes application like travel information, automated dialogue system, spoken language translation system, emergency assistance, language interpretation, buying services International markets and tourism add to the desirability of offering services in many languages. The languages of the world differ from one another along many dimensions which have been codified as linguistic categories. These include phoneme inventory, phoneme sequences, syllable structure, prosodic, phonotactics, lexical words and grammar. Therefore we hypothesize that an LID system which exploits each of these linguistic categories in turn will have the necessary discriminative power to provide good performance on short utterances.

————————————————

- *Pawan Kumar is with the Department of ECE, Birla Institute of Technology, Mesra, Ranchi, India.*
- *Astik Biswas is with the Department of ECE, Birla Institute of Technology, Mesra, Ranchi, India.*
- *A. N. Mishra is with the Department of ECE, Birla Institute of Technology, Mesra, Ranchi, India.*
- *Dr. Mahesh Chandra is with Birla Institute of Techonology, Mesra, Ranchi, India - 835215.*

A LID system has three major components, database preparation, feature extraction and classification as shown in Fig 1. A prerequisite for the development and evaluation of automatic speech recognition system is the availability of appropriate database. The recognition performance heavily depends on the performance of the feature extraction block. Thus choice of features and its extraction from the speech signal should be such that it gives high identification performance with reasonable amount of computation. There are two main methods used to parameterize the speech signal. Both PLP coefficients based on the linear prediction (LP) based technique [1, 2, 3] and MFCCs [4] based on discrete cosine transform (DCT) have some conceptual similarities of the speech signal processing. But there are some differences, which can be important for given conditions and for good identification performance.

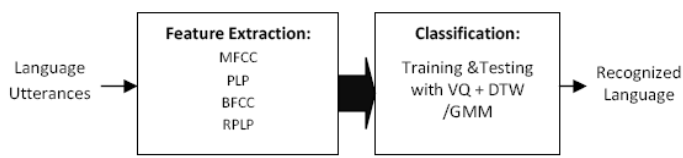

Fig 1:- Block diagram of Language identification system

The long time goal of the work is to use two hybrid robust feature extraction techniques which may give better identification performance for noisy environment as well as clean environment. Cepstral features were choosen because they yield high identification accuracy, and are invariant to fixed linear spectral distortion from recording and transmission environment. Since speech production is usually modeled as a convolution of the impulse response of the vocal tract filter with an excitation source, the cepstrum effectively de-convolves these two parts, resulting in a low time component corresponding to the vocal tract system and a high time component corresponding to excitation source. Two hybrid robust feature extraction techniques, revising PLP (RPLP)[5, 6] and Bark frequency cepstral coefficient (BFCC)[6] developed from basic parameteriza-tion methods PLP and MFCC were used as feature extraction techniques. Vector Quantization (VQ) [7, 8, 9], Dynamic Time Warping (DTW)[8, 10] and Gaussian Mixture Model (GMM)[11, 12] were used for classifying the languages into different classes. The Language Identification system was implemented with MATLAB 7.1.

This paper is organized in five sections. Section1 introduces the motivation of LID. Section2 gives the details of database preparation. Different feature extraction techniques are explained in section3. Experimental setup & result are given in section4. Finally the conclusions are drawn in section5.

## 2 DATABASE PREPARATION

A database for three Indian Languages (Bengali, Hindi and Telugu) has been prepared with 16 kHz sampling frequency with 16 bits resolution. Each language consists of seven different speakers and each speaker utterance was of one-minute duration. All speakers of respective Languages uttered same paragraph for one minute duration which were recorded in a noise-free environment. The foreign Language samples (Dutch, English, French, German, Italian, Russian, and Spanish) has been downloaded from Internet [13] and reformatted with 16 kHz sampling frequency & 16 bits resolution. Thus there were total ten Languages with seven different speakers for each language. So we have a total of 70 speech utterances. The duration of speech utterance of all languages ranges from 35 sec to 70 sec. Goldwave 5.10 & Cool Edit 96 software were used for database preparation and re-sampling.

## 3 FEATURE EXTRACTION

The raw speech signal is complex and may not be suitable for feeding as input to the automatic language identification system; hence the need for a good front-end arises. The task of this front-end is to extract all relevant acoustic information in a compact form compatible with the acoustic models. In other words, the preprocessing should remove all non-relevant information such as background noise and characteristics of the recording device, and encode the remaining (relevant) information in a compact set of features that can be given as input to the classifier. Features can be defined as a minimal unit, which distinguishes maximally close classes. The entire scheme for feature extraction using PLP, MFCC, BFCC and RPLP techniques are shown in Fig 2.

### 3.1 Mel Frequency Cepstral Coefficient

Pre-emphasis filtering, normalization and mean subtraction are the three steps in pre-processing. The digitized speech is pre-emphasized using a digital filter with a transfer function $H(Z) = 1 - 0.98z^{-1}$. Pre-emphasis filter [3] spectrally flattens the signal and makes it less susceptible to finite precision effects later in the signal processing. Due to possible mismatch between training and test conditions, it is considered good practice to reduce the amount of variation in the data that does not carry important speech information as much as possible. For instance, differences in loudness between recordings are irrelevant for recognition. For reduction of such irrelevant sources of variation, normalization transforms are applied.

During normalization every sample value of the speech signal is divided by the highest amplitude sample value. Mean of the speech signal is subtracted from the speech signal to remove the DC offset and some of the disturbances induced by the recording instruments. After pre-processing, language utterances were divided into different frames of 25ms duration. The second frame starts after 15ms of first frame and overlaps the first frame by 10ms. The third frame starts after 15ms of second frame and overlaps the second frame by 10ms and this process continue till the end of all the frames of speech sample. Next each frame was multiplied by a hamming window. After windowing first FFT and then Mel spaced filter banks are applied to get the Mel-spectrum. The Mel scale is logarithmic scale resembling the way that the human ear perceives sound.

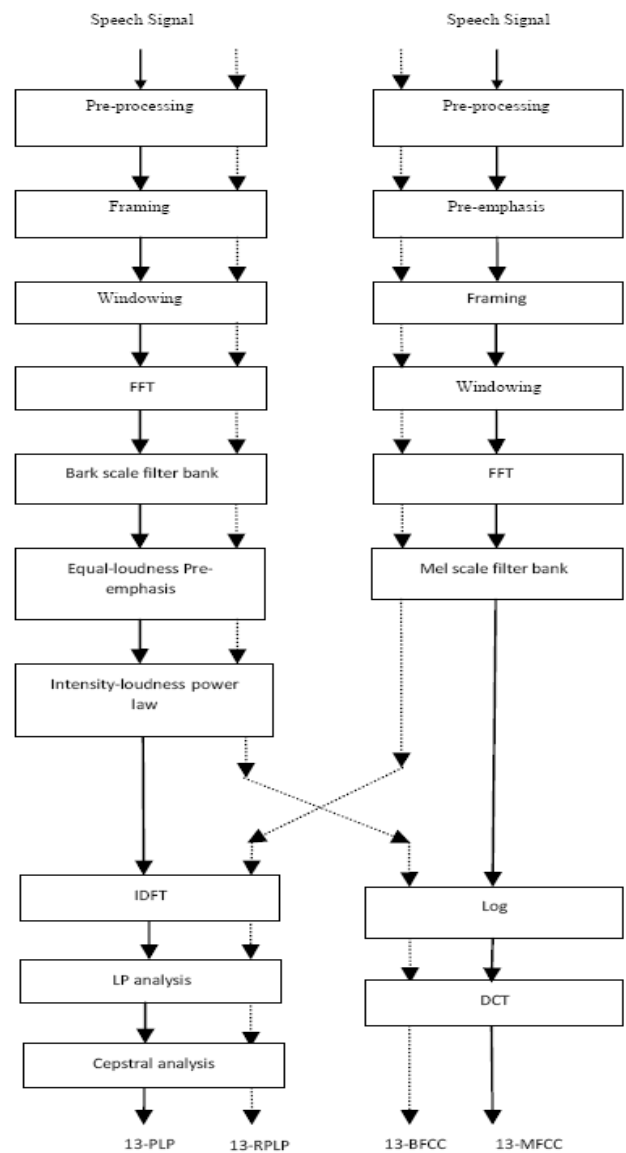

Fig 2: Feature extraction process for different methods

The filter bank is composed with 24 triangular filters that are equally spaced on a log scale. The Mel-scale is represented by

the following:

$$Mel(f) = 2595 \log_{10}\left(1 + \frac{f}{700}\right) \qquad \text{............... (1)}$$

The natural logarithm is taken to transform into the cepstral domain and the discrete cosine transform (DCT) is finally applied to get the 24 MFCCs. The component due to the periodic excitation source may be removed from the signal by simply discarding the higher order coefficients. DCT de-correlates the features and arranges them in descending order of information, they contain about speech signal. Hence 13 coefficients out of 24 coefficients are used as MFCC features in our case. MFCC features are more compact since the same information can be contained in fewer coefficients.

### 3.2 Perceptual Linear Prediction

An approach for linear prediction completely based on perceptual criteria is the Perceptual Linear Prediction (PLP) [5]. This model includes the following perceptually motivated analyses:

1. Critical-band spectral resolution. The spectrum of the original signal is warped into the Bark frequency scale, where a critical-band masking curve is convolved to the signal. For the case of PLP, trapezoidal shaped filters are applied at roughly one bark intervals, where the bark axis is derived from the frequency axis by using a warping function from Schroeder is given in equation number 2:

$$\varsigma(\omega) = 6\ln\left\{\frac{\omega}{1200\pi} + \left[\left(\frac{\omega}{1200\pi}\right)^2 + 1\right]^{0.5}\right\} \qquad \text{............ (2)}$$

2. Equal-loudness pre-emphasis. The signal is pre-emphasized by a simulated equal-loudness curve to match the frequency magnitude response of the ear.

3. Intensity-loudness power law. The signal amplitude is compressed by the cubic-root to match the nonlinear relation between intensity of sound and perceived loudness.

After these operations, all signal components are perceptually equally weighted and we can, from the modified signal, make a regular Linear prediction (LP) model.

### 3.3 Hybrid Features

In this experiment two main blocks as shown in fig.2 were interchanged to develop two hybrid feature extraction techniques. The interest is to see the influence of the spectral processing on the different cepstral transformation. The fig.2 shows the steps of parameterization for the basic method and besides PLP and MFCC the way of computing the hybrid techniques has been shown by dashed arrow in fig 2.

**Bark Frequency Cepstral Coefficient (BFCC)**

BFCC is the process where we combine PLP processing of the spectra and cosine transform to get the cepstral coefficients. Instead of using Mel filter bank, Bark filter bank has been applied and equal loudness pre-emphasis with intensity to loudness power law has been applied to the MFCC like features. Only first thirteen cepstral features of each windowed frame of speech utterances were taken.

**Revised Perceptual Linear Prediction (RPLP)**

In the second approach instead of using bark filter bank, Mel filter bank has been applied to compute RPLP. The signal is pre-emphasized before the segmentation and FFT spectrum is processed by Mel scale filter bank. The resulting spectrum is converted to the cepstral coefficients using LP analysis with prediction order of 13 followed by cepstral analysis.

## 4 EXPERIMENTAL SETUP & RESULTS

The whole work was carried in two different ways. In first phase, the work was carried out with all ten language database and VQ was used for creating the language model and DTW was used as classifier to classify languages into different classes. In second phase, the work was carried out with ten language database and Gaussian Mixture Model was used to generate the language models and GMM was used as classifier to classify languages into different classes.

### 4.1 Training and Testing (VQ + DTW)

Here VQ was used for training the language model and DTW was used as classifier to classify languages into different classes. The experimental set-up is shown in fig 3. By using MFCC, BFCC, PLP and RPLP techniques, features were extracted for each utterance of all 70 speakers of all languages.

The sub frames were of 25 ms with 10 ms overlapping. For all kind of feature extraction techniques thirteen features are calculated for each frame.

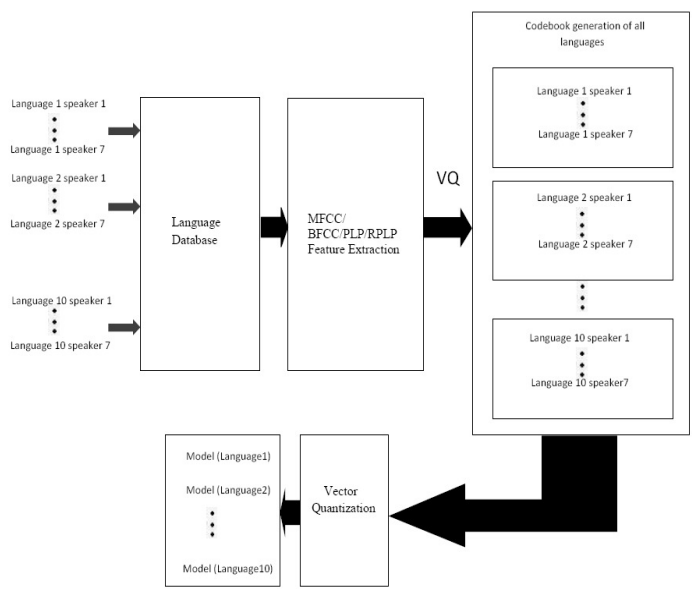

Fig 3: VQ codebook generation for all ten Languages

All feature vectors of all frames of an utterance were coded into single feature vector using VQ. In this way 70 feature vectors were prepared for all speakers of all languages. These feature vectors were stored for further use during classification. Further the seven feature vectors of all seven speakers of each language were coded into a single feature vector using VQ. Finally a total of 10 feature vectors were received, one feature vector corresponding to one language.

During testing phase, languages were classified into their respective classes by measuring similarity of each feature vector (of all 70 stored feature vectors) with the finally received ten feature vectors of ten languages. DTW was used for calculation to find similarity between two sequences which may vary in time. Comparative results for all the feature extraction techniques MFCC, BFCC, PLP and RPLP with VQ & DTW are shown in fig 4.

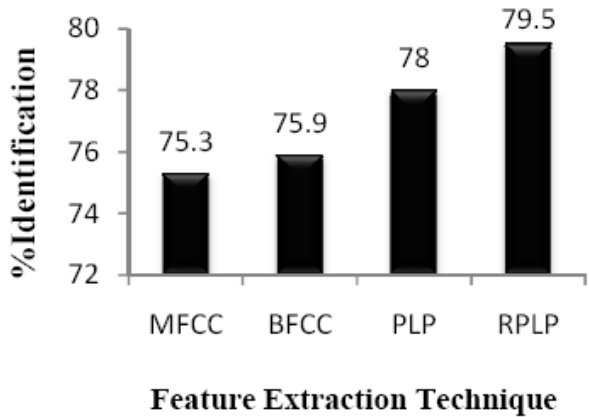

Fig 4: Language identification results with MFCC, BFCC, PLP and RPLP with VQ and DTW

From the result it can be found that the identification performance of BFCC is 0.6% better than MFCC features for LID. The identification performance with PLP is further improved by 2.1% over BFCC. RPLP has shown the best identification performance among all the feature extraction techniques.

### 4.1 Training and Testing (GMM+GMM)

During this phase, work was carried out by taking duration of thirty second of speech for each sample of each language. Here database was divided into two parts. First six utterances of each language were used for training purpose and last utterance was used for testing phase. During training phase, total of (6*30 = 180) seconds of speech per language was used to create one language model with 2, 4, 8, and 16 component densities. Finally total of 60 speech utterances were used for training and 10 speech utterances were used for testing. By using MFCC, BFCC, PLP and RPLP techniques, features were extracted for all 60 utterances of all languages. In order to have more temporal information, the duration of each sample was divided into number of sub-frames. The sub frames were of 25 ms with 10 ms overlapping. For all kind of feature extraction techniques thirteen features are calculated for each frame.

All feature vectors of all frames of all utterances were stored for further use in training. Then for all languages (i.e. feature vectors of first six utterance of that particular language) were used to create the corresponding language model with 2, 4, 8, and 16 component densities. In this manner all language models were created with 2, 4, 8, and 16 component densities [11].

The testing phase was divided into three different sub phase. At first testing was carried out for two second test utterances, then testing was carried out for four second test utterances and at last testing was carried out for ten second test utterances. The languages were classified into their respective classes on the basis of maximum log-likelihood [11, 12] w.r.t each language model The objective is to find the language model which has the maximum posteriori probability [14] for a given observation sequence. The whole experimental setup is shown in fig 5 and fig 6 respectively.

Comparative results for all the feature extraction techniques MFCC, BFCC, PLP and RPLP with GMM are shown in table 1.

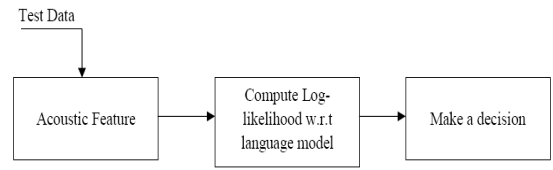

Fig 6: Testing Phase

Table1: Language identification performance (%) of MFCC, BFCC, PLP and RPLP features with GMM

| Feature Extraction Technique | 2 sec test utterances Number of Mixture Component | | | | | 4 sec test utterances Number of Mixture Component | | | | | 10 sec test utterances Number of Mixture Component | | | | |
|---|---|---|---|---|---|---|---|---|---|---|---|---|---|---|---|
| | 2 | 4 | 8 | 16 | Avg | 2 | 4 | 8 | 16 | Avg | 2 | 4 | 8 | 16 | Avg |
| MFCC | 75 | 77 | 78 | 81 | 77.25 | 70 | 78 | 82 | 80 | 77.5 | 76 | 80 | 88 | 78 | 80.5 |
| BFCC | 78 | 79 | 80 | 78 | 78.75 | 72 | 79 | 82 | 83 | 79 | 72 | 84 | 86 | 83 | 81.25 |
| PLP | 79 | 92 | 88 | 84 | 85.7 | 82 | 88 | 90 | 85 | 86.2 | 80 | 88 | 92 | 90 | 87.5 |
| RPLP | 84 | 89 | 90 | 86 | 87 | 80 | 85 | 93 | 86 | 86 | 84 | 92 | 90 | 89 | 88.75 |

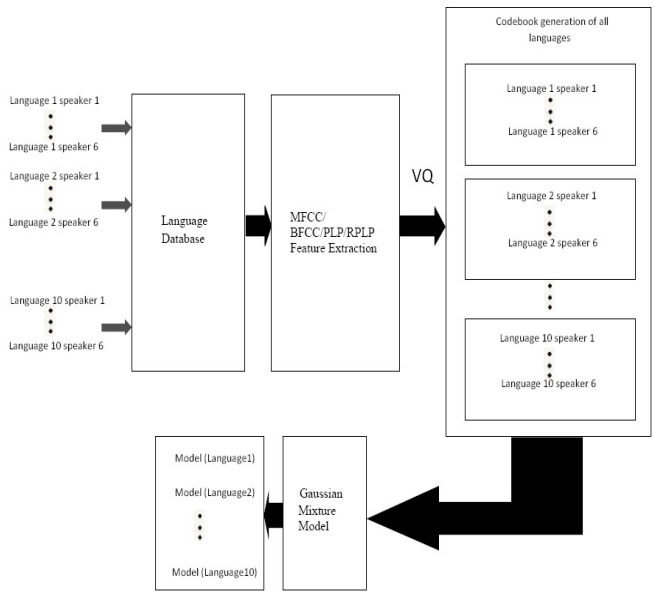

Fig 5: Mixture model generation for all ten Languages

Language identification results reveal that performace increases with increament in length of test utterances. For each model order, the identification performance for 2, 4, and 10 second test utterance lengths are shown in table1. BFCC has shown better identification performance compared to MFCC. Further the performance with PLP has increased compared to BFCC. In this case also RPLP has shown best identification performance among all. It also has been seen that in most of the cases performance peaks at 8 mixture components.

## 5 CONCLUSION

The Language Identification system provides satisfactory results by using four different types of features, MFCC, BFCC, PLP and RPLP with two classifiers, VQ along with DTW and GMM. BFCC has shown better identification performance compared to MFCC because it is more invariant to fixed spectral distortion and channel noise compared to MFCC. PLP features have shown further improvement in identification performance. This is due to the fact that PLP is combination of both MFCC and LP based features. PLP features performed better because the signal was pre-emphasized by a simulated equal-loudness curve to match the frequency magnitude response of the ear as well as all signal components were perceptually equally weighted. RPLP features performed best among all feature extraction techniques. This is due to the fact that it takes advantage of preemphasis filter, Mel scale filter bank along with Linear Prediction and cepstral analysis.

All feature extractiontechniques performed better with GMM as compared to VQ and DTW because gaussian mixture language model falls into the implicit segmentation approach to language identification. It also provides a probabilistic model of the underlying sounds of a person's voice.

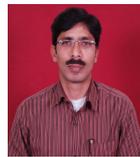
**Mr. Pawan Kumar** has received his M. Sc in 2005 from Ranchi University, Ranchi, India. Presently he is pursuing Ph.D. from Birla Institute of Technology, Mesra (Jharkhand)-India in the field of speech Recognition. His areas of interest are Speech and Signal Processing.

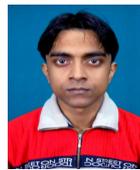
**Mr. Astik Biswas** has received his B.Tech in 2008 from West Bengal University of Technology, Kolkata, India. Presently he is pursuing ME from Birla Institute of Technology, Mesra (Jharkhand)-India in the field of speech Recognition. His areas of interest are Speech and Signal Processing, Digital Electronics.

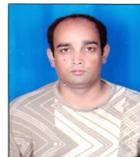
**Mr. A. N. Mishra** has received his B.Tech from Gulbarga University, Gulbarga- India in 2000 and M.Tech from Uttar Pradesh Technical University, Lucknow (UP)-India in 2006. Presently he is pursuing Ph.D. from Birla Institute of Technology, Mesra (Jharkhand)-India. He has worked as lecturer in the Department of Electronics & Communication Engg. at BBIET, Bulandshahr (UP) from Nov 2000 to July 2003. He has worked as Lecturer in the Department of Electronics & Communication Engg. at GLAITM ,Mathura (UP) from Oct 2003 to Aug 2005. He has worked as Reader in the Department of Electronics & Communication Engg. at BBIET ,Bulandshar (UP) from Aug 2005 to Oct 2007. He has worked as Reader in the Department of Electronics & Communication Engg. at RKGIT ,Ghaziabad (UP) from Oct 2007 to Jan 2009. Since Jan 2009, he is working as Reader and HOD in the Electronics & Communication Engg. Department, GGIT, Greater Noida, (UP)-India. He has published more than 4 research papers in the area of Speech, Signal and Image Processing at National/International level. His areas of interest are Speech, Signal and Image Processing.

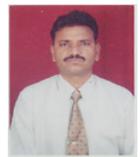
**Dr. Mahesh Chandra** received B.Sc. from Agra University, Agra(U.P.)-India in 1990 and A.M.I.E. from I.E.I., Kolkatta(W.B.)-India in winter 1994. He received M.Tech. from J.N.T.U., Hyderabad-India in 2000 and Ph.D. from AMU, Aligarh (U.P.)-India in 2008. He has worked as Reader & HOD in the Department of Electronics & Communication Engg. at S.R.M.S. College of Engineering and Technology, Bareilly (U.P.)-India from Jan 2000 to June 2005. Since July 2005, he is working as Reader in the Electronics & Communication Engg. Department, B.I.T., Mesra, Ranchi (Jharkhand)-India. He is a Life Member of ISTE, New Delhi-India and Member of IEI Kolkata (W.B.)-India. He has published more than 23 research papers in the area of Speech, Signal and Image Processing at National/International level. He is currently guiding four Ph.D. students in the area of Speech, Signal and Image Processing. His areas of interest are Speech, Signal and Image Processing.